\documentclass[twoside,a4paper,11pt]{proceedings}
\usepackage{graphicx}
\usepackage{hyperref}
\usepackage{movie15}
\usepackage{natbib}
\begin{document}
\pagenumbering{arabic}
\pagestyle{myheadings}
\thispagestyle{empty}
\vspace*{-1cm}
\vspace*{0.2cm}
\begin{flushleft}
{\bf {\LARGE
Optical linear polarization measurements of WR massive binary and single stars
}\\
\vspace*{1cm}
Akras S.$^{1,2}$,
Ramirez--Velez J.$^2$,
Hiriart D.$^2$,
and
Lopez J. M.$^2$
%
}\\
\vspace*{0.5cm}
%
$^{1}$
Institute of Astronomy, Astrophysics, Space Applications \& Remote Sensing, NOA\\
$^{2}$
Instituto de Astronomia, Universidad Nacional Aut\'onoma de Mexico, Ensenada, Mexico
%
\end{flushleft}
\markboth{
Optical Polarimetry of Massive stars
}{
Akras S. et al.
}
\thispagestyle{empty}
\vspace*{0.4cm}
\begin{minipage}[l]{0.09\textwidth}
\ 
\end{minipage}
\begin{minipage}[r]{0.9\textwidth}
\vspace{1cm}
\section*{Abstract}{\small
We present optical (UBVRI) linear polarimetric observations of 8 Wolf--Rayet (WR) massive 
binaries and single stars. We have corrected the observed values for the interstellar extinction 
and polarization by the interstellar medium to obtain the intrinsic polarization and position angle. 
We find three highly polarization stars between 5\% and 10\% (WR1, WR5 and WR146), three between 3\% 
and 4\% (WR2, WR3 and WR4), and two between 1\% and 2\% (WR137 and WR140). Moreover, 5 stars show 
increasing degree of polarization to shorter wavelengths (e.g WR 146) indicative with asymmetric 
circumstellar envelope and 3 have nearly constant polarization within the errors (e.g WR 140).
\vspace{10mm}
\normalsize}
\end{minipage}

\section{Introduction}


Massive stars are of one of the most puzzling components of stellar evolution and many 
questions related to them still remain unanswered. They show high mass-loss rate (10$^{-6}$ to 10$^{-7}$ M$_{\odot}$ y$^{-1}$), 
in the form of winds or ejecta. Specifically, Wolf--Rayet (WR) stars represent a late evolutionary 
phase of massive stars with faster winds and significantly higher mass-loss rate (10$^{-4}$ to 10$^{-5}$ M$_{\odot}$ y$^{-1}$) 
than OB massive stars (Nugis \& Lamers 2000). 39\% of the Galactic WR stars may present a binary companion (van der Hucht 2001). Despite the fact that 
the intense radiation field of the young companion should prevent the formation of dust in these binaries, 
Allen, Swings \& Harvey (1972) found substantial IR excess in a subset of WR stars indicative with the 
presence of circumstellar dust at 1500 K, associated with strong colliding-winds. It is expected, therefore, 
to be highly polarized due to the free-free scattering (e.g. WR 97, ~5.5\%, Niemela et al. 1996). 
Nevertheless, Harries, Hillier \& Howarth (1998) did not found evidence of linear polarization among WR binaries. 
Polarimetry is, therefore, an important observational technique to study the dust formation and properties 
-(distribution, composition) and stellar winds of massive stars.

\section{Observations}

Optical (UBVRI) linear polarimetry of 8 massive stars were performed 
at the 0.84m, f/15 telescope at Observatorio Astronomico Nacional in Mexico 
using a CCD camera and the POLIMA polarimeter (Hiriart et al. 2005) in August 2012 and April 2013. 
The normalized Stokes parameters Q and U were calculated by mean of the following equations: 
{\it Q=f(0)-f(90)/f(0)+f(90) \& U=f(45)-f(135)/f(45)+f(135)} 
where f($\theta$) is the flux of the object in the image at a position angle $\theta$ of the polarizer optical 
axis. Q and U parameters are related to the degree (P) and position angle ($\theta$) of polarizations by the 
following equations: {\it P=(Q$^2$+U$^2$)1/2 \& $\theta$=arctan(U/Q)/2}.
Thus, we assumed that the circular polarization is neglectable.

\section{Preliminary Results}

\begin{figure}
\includegraphics[scale=0.31]{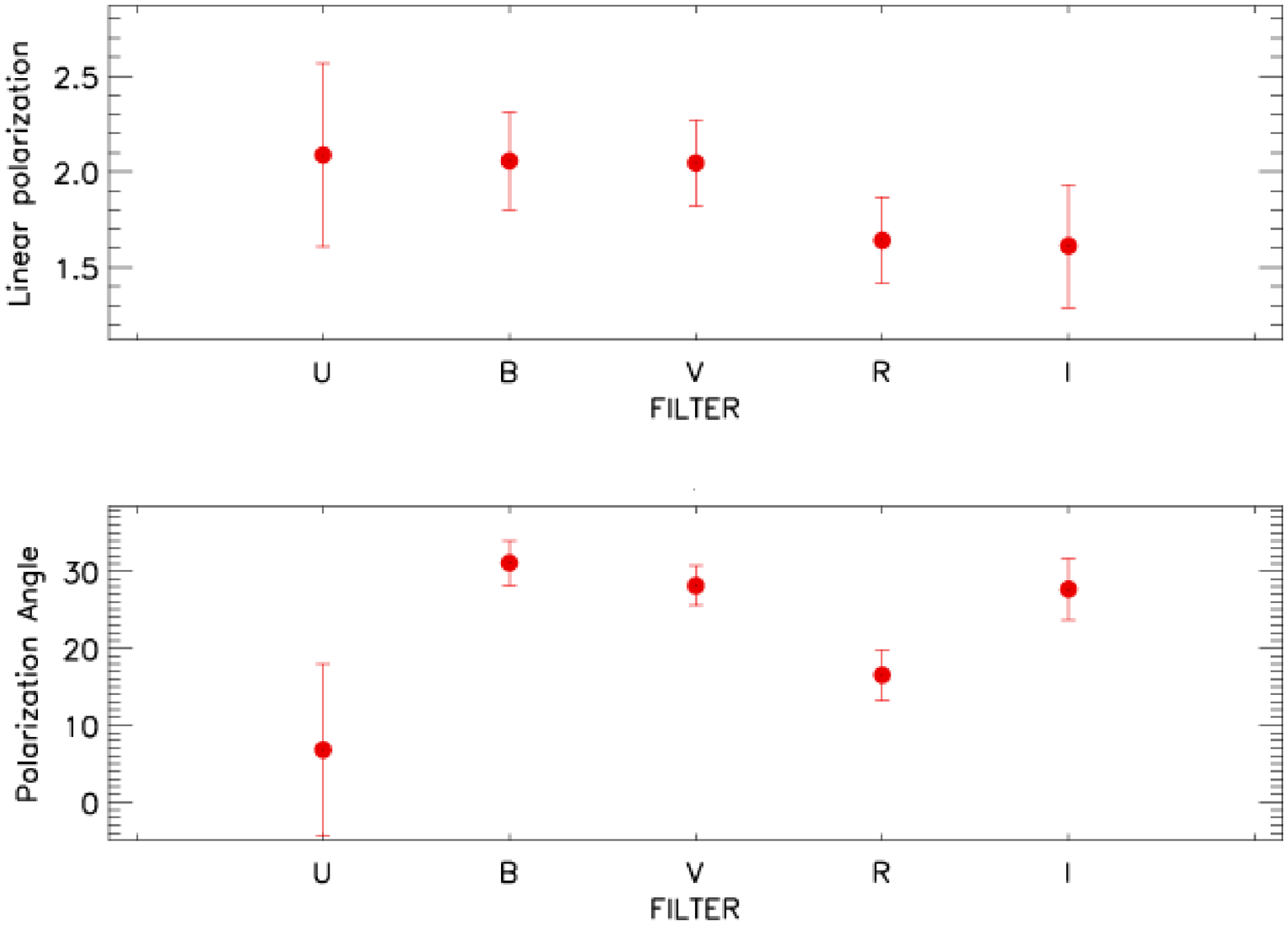} 
\includegraphics[scale=0.31]{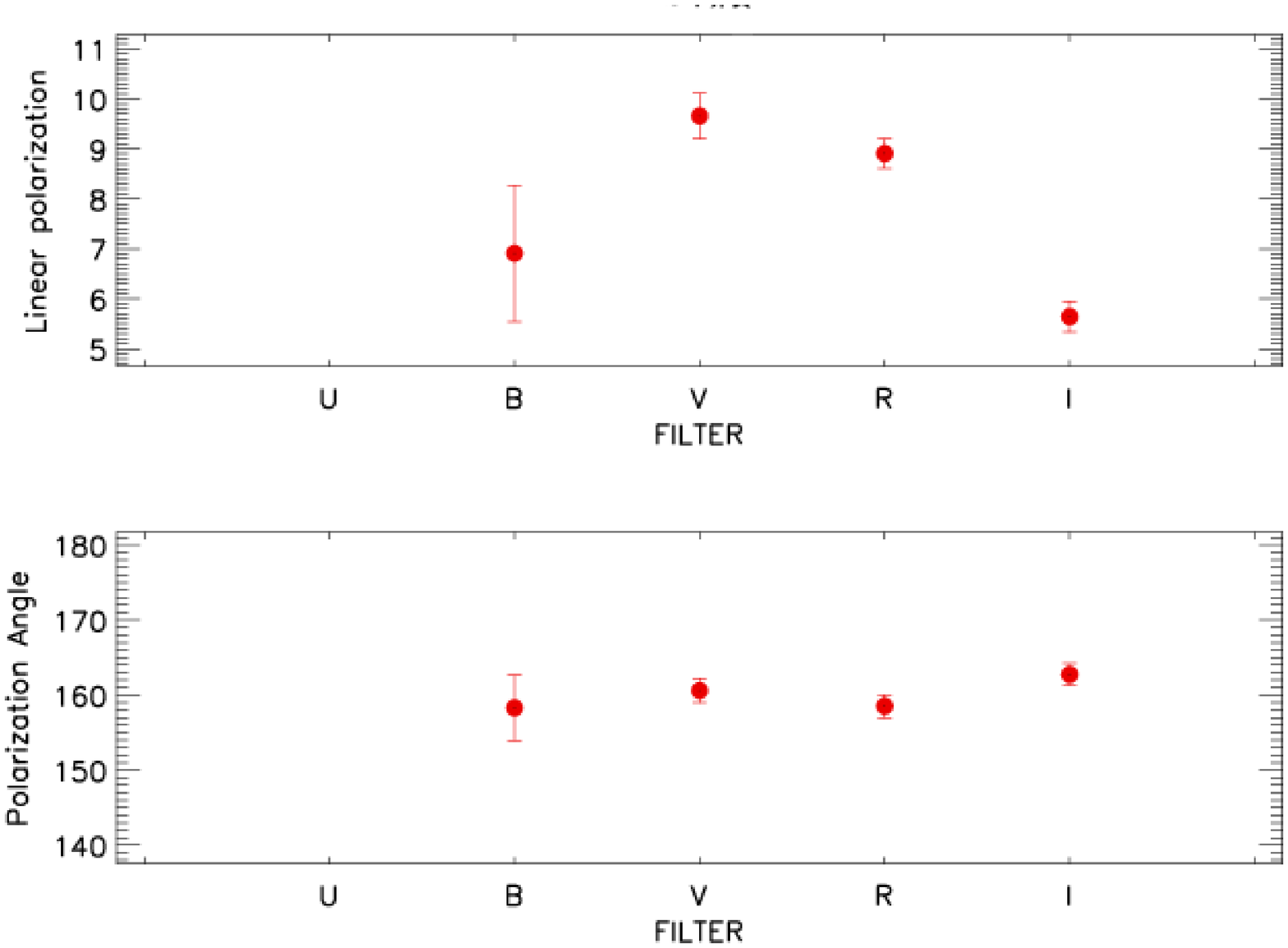} 
\caption{Polarization (upper panel) and position angle of the polarization vector (lower panel) 
as a function of wavelengths for WR140 (left panels) and WR146 (right panels).}
\end{figure}

Our prelimiary results suggest scattering of light from dust grains in axisymmetric circumstellar 
envelopes around Massive WR stars. WR140 is a well--known colliding wind binary star and periodic dust--maker (Williams et al. 1990). 
The degree of polarization is found to be independent of wavelength at about 2\% indicative or large dust grains scatters 
(Fig1. left panels). WR3 and WR137 show similar flat polarization wavelength dependence at 2.5\% and 1\%, respectively.
WR146 is found to be the highest polarized star in our sample with increasing polarization to 
shorter wavelengths (Fig1. right panels) consistent with an axymmetric circumstellar 
envelope and multiple scattering (Wood et al. 1996). WR1, WR2, WR3 and WR5 show similar 
wavelengths dependence and maximum polarization in the B-band of 6.5\%, 4.0\%, 3\% and 4.5\%, respectively.  

WR1 shows small time variations of polarization and position angle (Fig. 2). Responsible for these time 
variations may be: (i) asymmetric circumstellar/binary envelopes, (ii) episodic dust formation due 
to colliding winds, (iii) inhomogeneities in WR winds (e.g blobs) and (iv) the orbital phase for binary systems. 
However, low--level variations can also be occurred due to the weather conditions, calibration and technical problems.

\section{Conclusion}

Our main conclusion are: (i) 6 WR stars are clearly polarized. Three of them 
show intrinsic linear polarization higher than 5\%, three have polarization between 3\% and 
4\%, and two lower than 2\%, (ii) we did not find difference between WN and WC subclasses 
indicating no correlation between the polarization and Teff (evolutionary stage of WR stars), 
(iii) variable plarization might be found due to dust formation in colliding winds zone 
and asymmetric circumstellar/binary envelopes and (iv) linear polarization seems to increase 
with J--H color. Surprisingly, we did not find a similar trend for J-K and H-K colors.

\begin{figure}
\center
\includegraphics[scale=0.46]{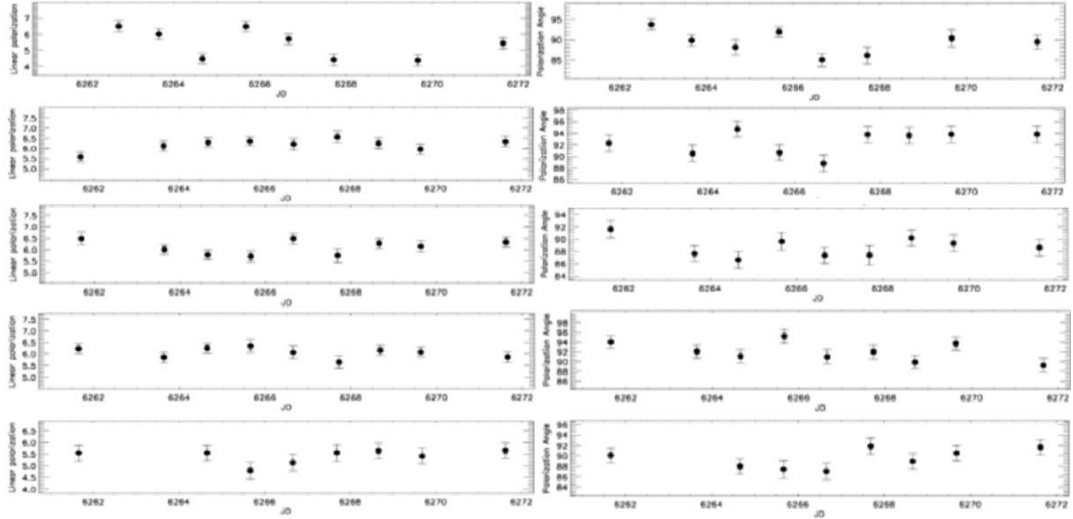} 
\caption{Polarization (left panels) and position angle (right panels) vs Julian Date for the WR1 in the 
U, B, V, R and I bands}
\end{figure}

\small  
%
\section*{Acknowledgments}   
%
S. A. gratefully acknowledges a postdoctoral scholarship from DGAPA--UNAM (IN100410). 
R.--V. J. and H. D. also acknowledge support from CONACyT, Mexico through grants 180817 and 106719.

\section*{References} 
\bibliographystyle{aj}
\small
Allen D. A., Swings J. P., \& Harvey P. M., 1972, A\&A, 20, 333 \\
Harries T. J., Hillier D. J., \& Howarth I. D., 1998, MNRAS, 296, 1072 \\
Hiriart D., Valdez J., Quiros F., et al., 2005, POLIMA Manual de Usuario, UNAM\\ 
Niemela V. S., Rovero A., C., \& Cerruti M. A., 1996, RMxAC, 5, 126 \\
Nugis T., \& Lamers H. J. G. L. M., A\&A, 2000, 360, 227 \\
van der Hucht K. A., 2001, New Astr. Rev. 45, 135 \\
Williams P. M., van der Hucht K. A., Pollock A. M. T., et al., 1990, MNRAS, 243, 662\\ 
Wood K., Bjorkman J. E., Whitney B., et al., 1996, ApJ, 461, 847
\bibliography{proceedings}

\end{document}